# Authentication against Man-in-the-Middle Attack with a Time-variant Reconfigurable Dual-LFSR-based Arbiter PUF

Yao Wang and Zhengtai Chang

*Abstract*—With the expansion of the Internet of Things industry, the information security of Internet of Things devices attracts much attention. Traditional encryption algorithms require sensitive information such as keys to be stored in memory, and also need the support of operating system, which is obviously unacceptable for resource-constrained Internet of Things terminals. Physical not cloning function by extracting the chip is inevitable in the process of manufacturing process deviation, the introduction of the corresponding function relationship between incentive and response, not to need the storage user sensitive information, and only when electricity will respond, in power response immediately disappear, this can save a lot of resources of equipment and the power consumption. However, PUF is vulnerable to modeling attacks, and the traditional methods such as the challenge obfuscation method are time-invariant, which is equivalent to adding a fixed function to the front stage of a traditional APUF circuit. Therefore, it can be potentially modelling attacked with sufficient CRPs. In order to further enhance APUF circuit resistance to modelling attack, this paper proposes a dual-LFSR-based APUF circuit with time-variant challenge obfuscation. Besides, traditional authentication scheme generally adopts the one-time key scheme to enhance resistance to man-in-the-middle attack. The two-time authentication scheme proposed in this paper can improve the ability of RFID system to resist man-in-the-middle attack without sacrificing CRPs.

## I. Introduction

With the rapid development of information technology, the Internet of Thing(IoT) industry has expanded rapidly, at present, the world has more than 10 billion devices are connected, and in the next five years is expected to deploy more than five times the device [1]. The market demand of Radio Frequency Identification (RFID) and other electronic device is becoming more and more big. While the development of Internet of Things technology brings great convenience to people's life, the information security of Internet of Things devices also attracts much attention. In the information society, information security is a problem we have to face.

At present, the most widely used equipment safe solution is to adopt the traditional software encryption algorithms, such as AES, RSA and digital signature [2], but the traditional software encryption algorithm is usually the user's sensitive information (such as key) block is stored in the EEPROM, FLASH [3] and other nonvolatile memory or storage in the power supply power for random access memory (RAM). As IoT devices are resource-limited hardware platforms, their volume and power consumption are strictly limited. However, traditional software encryption algorithms not only need to store sensitive information in memory, but also need the support of a larger operating system, which is obviously unacceptable for resource-limited IoT terminals [4]. In addition, traditional software encryption algorithms need to store sensitive information in memory, which is vulnerable to physical attacks such as side channel attack, intrusion/semi-intrusion attack [5]. Traditional software encryption algorithms are quite vulnerable to such attacks.

Physical Unclonable Function (PUF) takes advantage of the inevitable random differences in chip making, extracts them and presents them as binary sequences of signals [6]. In the manufacturing process of an integrated circuit, even if two chips are identical in structure and so on, there must be a small difference in internal delay and so on. This difference is also uncontrollable to the manufacturer, so PUF circuit cannot be cloned. And PUF circuit respond only when the electricity, the electrical response immediately disappear, which not only overcomes the traditional software encryption algorithms rely on storage key faults, also greatly improve the ability against the attack of the equipment, reduces the equipment and power consumption of resources, is very suitable for the Internet of things terminal information encryption.

However, traditional PUF circuits are seriously threatened by modeling attacks. The attacker collects a large number of CRPs of PUF circuit, uses machine learning algorithm to build a model, and then predicts undiscovered CRPs of PUF circuit. With the deepening of the research on PUF structure, the prediction rate of modeling attack on traditional PUF circuit is getting higher and higher. As a strong PUF, APUF is widely used in identity authentication due to its large CRPs. However, due to the inherent linearity of APUF circuits, conventional APUF is relatively easy to be breached in the face of machine learning attacks [7]. At present, there are several methods to improve APUF's resistance to machine learning attacks, one of which is the challenge obfuscation method [8]. The pros and cons are as follows.

Pros: Without changing the underlying architecture of APUF,

the external challenge of APUF is obfuscated to break the linear relationship between the traditional challenge and response.

Cons: The current challenge obfuscation methods for APUF are basically time-invariant, which is equivalent to adding a fixed function to the front stage of a traditional APUF circuit. Therefore, it can be potentially modelling attacked with sufficient CRPs.

In order to further enhance APUF circuit resistance to modelling attack, this paper proposes a dual-LFSR-based APUF circuit with time-variant challenge obfuscation.

In addition, in order to resist the man-in-the-middle attack, the traditional authentication scheme generally adopts the one-time key scheme. Once CRP is applied, it is immediately discarded. The scheme also sacrifices a large amount of CRPs resources while enhancing the resistance of RFID systems to attack. In this paper, combined with the proposed dual LFSR APUF structure, the two-time authentication mechanism can enhance the man-in-the-middle attack of RFID systems without sacrificing CRPs resources.

## II. PROPOSED PUF DESIGN

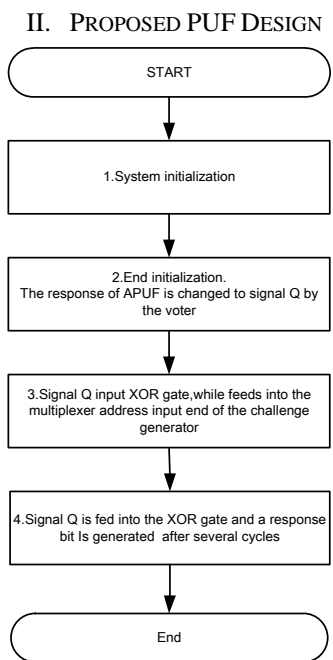

Fig. 1. The design flow for dual LFSR arbiter PUF

This section presents a focus on a variety of measures, such as randomness and reliability design method for dual-LFSR arbiter PUF, we first discuss how to improve randomness of APUF, avoid the path up and low in the FPGA chip to generate serious deviation, we also discussed how to improve the reliability of APUF to makes it to be implement on the given 45 nm FPGA chip. Moreover, when the FPGA platform is changed, it can be adjusted adapatively to improve its randomness and other measures. Finally, after the architecture of the underlying APUF is determined, the principle of dual LFSR is discussed, and it is shown that dual LFSR architecture can obscure challenges, thus improving its ability to resist machine learning attacks.

### A. Design method of dual LFSR arbiter PUF

The design flow shown in Fig. 1 can be used to design the dual-LFSR APUF. On the whole, the dual-LFSR APUF is divided into two parts: the top circuit and the bottom circuit. Bottom circuit is a basic APUF, and its design focuses on the realization platform of APUF circuit, that is, it is implemented on FPGA board. APUF circuit based on FPGA board will face serious problems of randomness and reliability. Therefore, when we design the bottom APUF circuit, we need to improve its randomness and reliability to facilitate the operation of the top circuit, which is also beneficial to the realization of the authentication protocol against man-in-the-middle attack.

In the process of FPGA automatic layout and routing, APUF's upper and lower paths will not be completely symmetrical, which will introduce certain layout and routing deviation. In order to eliminate the deviation as much as possible, the Randomness Adjustment Module is introduced, which mainly works in the System initialization stage. When the RST key on the FPGA board is pressed, the system enters the System initialization stage. The detailed working process of Randomness Adjustment Module is described in Section II-B.

After the signal passes through the Randomness Adjustment Module and the Voter Module, the randomness and reliability of the responses generated by APUF get better. In order to make APUF more resistant to attack, meantime, the proposed APUF can be used for authentication against man-in-the-middle attack. The dual LFSR structure is introduced, which generates the real challenge signal by two LFSRs jumping to each other with certain rules. This mechanism of generating real challenges can effectively enhance the resistance to machine learning attack of APUF by obfuscating real challenges. The detailed working process of dual-LFSR is described in Section II-C.

After System initialization, the path deviation caused by the asymmetry of APUF's upper and lower delay paths is basically eliminated. In order to improve the reliability of APUF circuit, the Voter Module is introduced, and the principle of minority subordinate to majority is adopted to improve the reliability of APUF at the cost of time redundancy. The analysis shows that the greater the time redundancy is, the stronger the reliability will be. Therefore, it is necessary to make a tradeoff between reliability and redundancy better when implementing in combination with specific scenarios. Furthermore, in order to further obscure challenge response pair, the output of the former Voter Module needs to go through the XOR gate, and after several cycles, the outputs of the XOR gate inputs the D flip-flop as a bit response of the PUF circuit. The detailed working process of Voter Module and subsequent module is Section II-D.

*B. Design of Randomness Adjustment Module*

When the RST key on the FPGA board is pressed, the PUF circuit enters the System initialization stage. The Randomness Adjustment Module begin to count the pulse signal of the input to APUF, and at the same time begin to count the "0" of the output of the arbiter (D flip-flop). After a certain number of pulses enter the APUF, the proportion of "0" in the response bitstream of the arbiter is calculated. If the proportion is close to 50%, it indicates that delay of the upper and lower two paths of APUF is basically same delay, namely, the randomness of APUF is relatively good. On the country, if the proportion is far from 50%, it indicates that deviation of the upper path and lower path of APUF is relatively large, and the randomness needs to be adjusted by compensating for the shorter delay path so that the upper and lower paths are symmetric.

The principle of compensation is successive approximation, that is, a small fixed delay amount is added to the shorter delay path, and then the proportion of "0" in the response bit of the arbiter is calculated again. If it is still far from 50%, the compensation continues until the proportion is close to 50%.

At the initial stage of the system, there are two important parameters in the Randomness Adjustment Module:

1) The number of pulse signals input to APUF. Obviously, the more the number of pulses counted, the less contingency there will be in statistical data, and the randomness index calculated from these data is closer to the true value of APUF circuit. However, the number of pulse signals input to APUF should not be too large, which will increase the initialization time of the system and reduce its working efficiency.

2) Threshold that determines if randomness reaches target, that is, the difference value between the proportion of "0" in the response bitstream and 50% can be considered as meeting the requirement. Obviously, the smaller the difference is, the better the randomness index after adjusting APUF (close to 50% of the ideal value). However, if the difference value is too small, that is, the randomness requirement of APUF is too high, the randomness adjustment module may need to go through lots of rounds of adjustment to meet the requirements, which will also increase the System initialization time and reduce its working efficiency. Through the analysis of 1) and 2), from the perspective of improving the performance of APUF system, we should increase the number of pulse signals and reduce the difference value between threshold and 50%. However, this will increase the time redundancy of the system. Therefore, in practice, we need to set these two parameters in combination with the specific scene.

The principle of Randomness Adjustment Module is shown in the diagram in Fig. 2, The meanings of each symbol in the diagram are as Table. I.

Here, we set the number of pulses in the input APUF to be 96, so in an ideal situation, "0" in the response bit should account for 50%, i.e., 48. Meantime, we set the difference between the threshold value and the ideal value as 6, that is when the number of "0" in the response bitstream is between 42 and 54, we believe that the APUF circuit meets the randomness requirement.

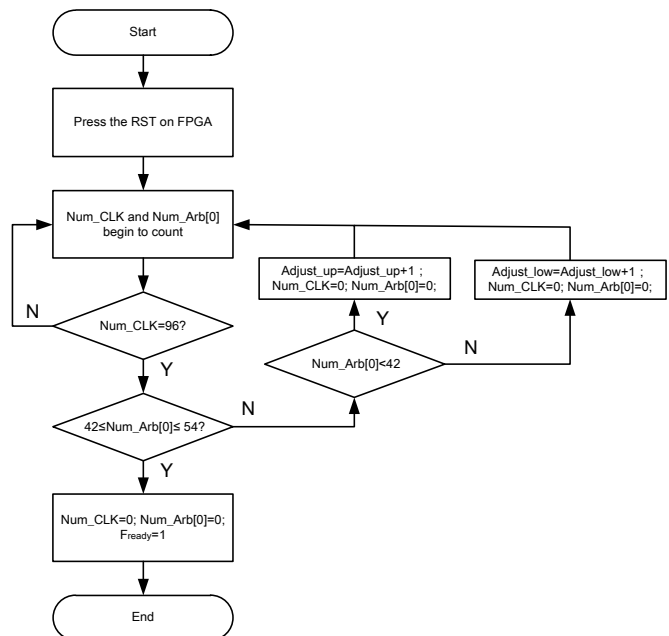

Fig. 2. The working flow of Randomness Adjustment Module

From Fig. 2, we can see that when Num_CLK counts to 96, we need to determine whether Num_Arb[0] is greater than 42 and less than 54. If so, we find the balance point, then we set Num_CLK and Num_Arb[0] to 0, note that Adjust_up and Adjust_low cannot be set to zero, but should remain unchanged. Meantime, we set the flag bit $F_{ready}$ to "1" for use by other modules at the same time, indicating that the adjustment of the Randomness Adjustment Module has ended, namely completion of System initialization. If Num_Arb[0] is less than 42, it means that the number of "1" in the response bitstream is too large, and the upper delay path needs to be increased, that is, a compensation unit is added in the upper compensation path. Then Num_CLK and Num_Arb[0] should set to 0 and begin to count again. Conversely, if Num_Arb[0] is greater than 54, it means that the number of "0" in the response bitstream is too large, a compensation unit is added in the lower compensation path. As before, Num_CLK and Num_Arb[0] should set to 0 and begin to count again.

TABLE I
MEANINGS OF EACH SYMBOL IN THE DIAGRAM

| Symbol | Meaning |
|---|---|
| Num_CLK | pulse number input to APUF |
| Num_Arb[0] | number of "0" in the output response bitstream |
| Adjust_up | number of compensation units in the upper compensation path in the Randomness Adjustment Module. If the upper delay path is relatively short in each round of judgment, the variable adds 1 |
| Adjust_low | number of compensation units in the lower compensation path in the Randomness Adjustment Module. If the lower delay path is relatively short in each round of judgment, the variable adds 1 |
| $F_{ready}$ | The flag bit that indicates the completion of System initialization. If the flag is 1, system initialization has been completed; otherwise, system initialization is in progress |

## C. Design of dual-LFSR structure

The linear feedback shift register takes the linear function of the output of the previous state and inputs it again into the shift register, the initial value of the input to LFSR is called the seed. LFSR is simple in structure and easy to implement. It can be used to produce uniformly distributed pseudo-random sequences. The n-order linear feedback shift register consists of n flip-flops and several XOR gates. According to the connection mode between XOR gate and flip-flop, linear feedback shift register can be divided into two realization modes, namely Fibonacci LFSR and Galois LFSR [9]. Since Fibonacci LFSR's XOR gates are all on the feedback loop, the combined logic delay of the feedback loop will increase. In order to further improve the circuit speed of LFSR, Galois LFSR is selected here. The n-order LFSR structure implemented by Galois is shown in Fig. 3.

The period of LFSR is only related to its feedback mode and has nothing to do with its seed value. According to the different feedback mode, the characteristic polynomial of LFSR is derived. The characteristic polynomial of LFSR is a digital model that represents the structural characteristics of linear shift registers. The characteristic polynomial of Galois LFSR shown in Fig. 3 is shown in Equation (1),

$$p(x) = g_0 + g_1 x + g_2 x^2 + g_3 x^3 + \cdots + g_n x^n$$

Where $g_i$ is 0 or 1, if $g_i$ is 0, there is no feedback at that point, if $g_i$ is 1, there is feedback at that point. Note that in the n order LFSR, $g_n$ is always assumed to be 1. Otherwise, if $g_n$ is 0, the n order LFSR will degenerate to (n-1) order LFSR. Always assume that $g_0$ is 1, otherwise if $g_0$ is 0, there is no feedback loop and the linear feedback shift register becomes a linear shift register. n order LFSR has at most $2^n$ states, but under linear operation, all 0 states will not be transferred to other states, so LFSR traverses at most $2^n-1$ states, that is, the longest period of output sequence is $2^n-1$, we call such sequence as the maximum length sequence, also known as m-sequence. In order to distinguish the sequence types (state types) of LFSR, we divide all sequences of LFSR into useful sequences, useless sequences and additional sequences. Among them, all 0 sequences are called useless sequences, the main cyclic sequences in the state transition diagram are called useful sequences, and the remaining sequences are called additional sequences. To further illustrate the three kinds of sequences, LFSR of order 3, whose characteristic polynomial is $x^3+x^2+x+1$, is taken as an example. The classification of all sequences is shown in Fig. 4. Note that the LFSR's useful sequence cannot contain the repetition sequence of a certain sequence; not all LFSRs have additional sequences. For LFSRs that can produce m-sequences, the number of useful sequences is up to $2^n-1$. Therefore, all sequences can only be divided into useful sequences and useless sequences for LFSR with m-sequences. In addition, in practical application, when LFSR is used as a pseudo-random number generator, the pseudo-random sequence we need is the useful sequences generated by LFSR.

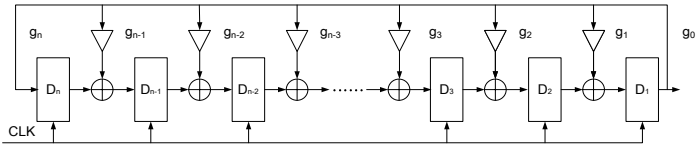

Fig. 3. Galois LFSR structure

The proposed PUF obfuscates challenge by dual-LFSR structure, the schematic is shown in Fig.5. The external challenges(green challenges) are the seeds of both LFSR. The real challenges of the underlying APUF circuit (black challenges) may come from LFSR1 (blue challenges) or LFSR2 (red challenges). Specifically, the real challenges of the underlying APUF circuit are determined by the previous response bit of the APUF. If the previous response bit of APUF is "0", the current real challenges are derived from LFSR1. If the previous response bit is "1", the current real challenges are derived from LFSR2. The challenge bit of LFSR1 is $C_{1i}(k)$, the challenge bit of LFSR2 is $C_{2i}(k)$, and the current real challenge bit is $\widehat{C}_i(k)$. The relationship between the current real challenge $\widehat{C}_i(k)$ and the previous response bit $R(k-1)$ is shown the following.

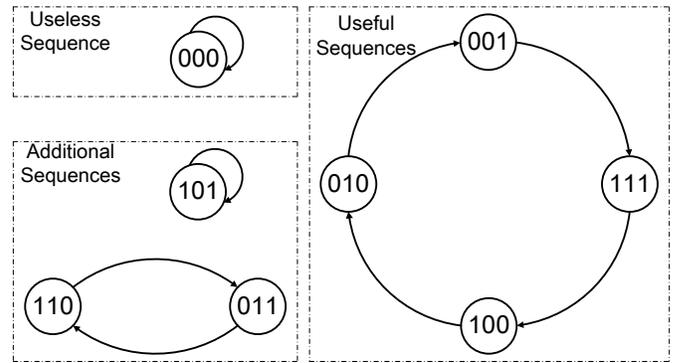

Fig. 4. Classification of all sequences of 3-order LFSR with characteristic polynomial $x^3+x^2+x+1$

$$\widehat{C}_i(k) = \begin{cases} C_{1i}(k) & R(k-1) = 0 \\ C_{2i}(k), & R(k-1) = 1 \end{cases}$$

Different from the obfuscation challenge of the previous time-invariant mapping function, this obfuscation challenge method is essentially a time-varying mapping function. In addition to the obfuscation challenge, a time factor and the previous response bit factor are introduced into the mapping function, as shown in Equation (2),

$$\hat{C}(k) = \hat{f}(C(k), t, R(k-1))$$

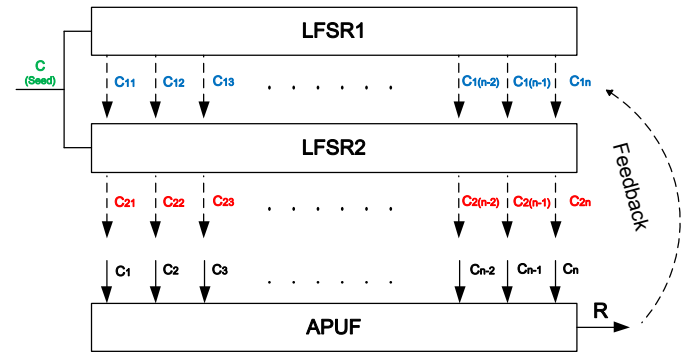

Fig. 5. Schematic of obfuscation challenge for dual-LFSR structure

Where $\hat{C}(k)$ is current real challenge which is actually the seed of two LFSRs, $C(k)$ is current external challenge, $t$ is the time interval between the input of the current external challenge $C(k)$ and the present (number of clocks), $R(k-1)$ is previous response bit of APUF. Compared to the previous time-invariant function mapping $f$ of a single variable, this time-variant mapping function $\hat{f}$ of multiple variables is more complex and difficult to predict.

In the dual LFSR structure proposed in this paper, we need to construct two LFSR of the same order with the maximum period and the same useful sequences. After analysis, we need to obtain two groups of m-sequences of the same order LFSR, which can meet the above requirements. If an n-order LFSR produces an m-sequence with period number $2^n-1$, then its characteristic polynomial is irreducible, but not all irreducible polynomials can generate m-sequence, and the irreducible polynomial that can generate m-sequence is called the primitive polynomial. We can use Matlab software to solve the primitive polynomial.

In order to quantitatively explain the working principle of APUF with dual LFSR structure, we take the 3-order LFSR as an example for analysis. First of all, there are two primitive polynomials of 3-order LFSR, which are $x^3+x+1$ and $x^3+x^2+1$ respectively. According to both primitive polynomials, the corresponding Galois LFSR is shown in Fig. 6 and Fig. 7 respectively. For the 3-order Galois LFSR in Fig. 6, it is assumed that the seed value at the beginning is $(D_3D_2D_1)=$

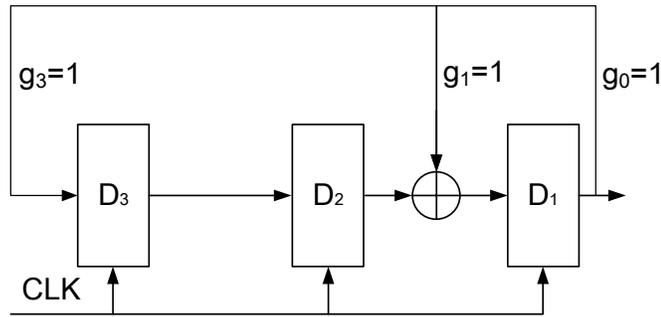
Fig. 6.  3-order Galois LFSR with characteristic polynomial $x^3+x+1$

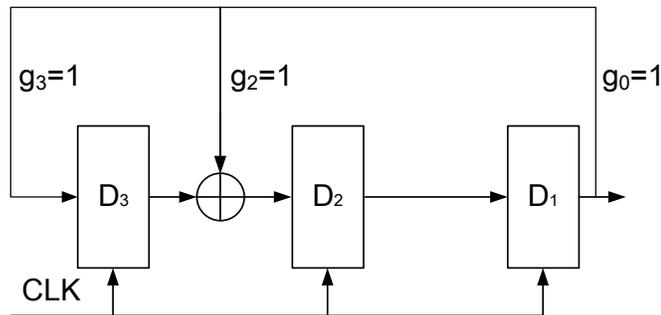
Fig. 7.  3-order Galois LFSR with characteristic polynomial $x^3+x^2+1$

(001), according to the state transition equation:

$D_3 = D_1$
$D_2 = D_3$
$D_1 = D_2 \wedge D_1$

It can be concluded that when the next clock edge arrives, the output value of LFSR jumps, $(D_3D_2D_1) = (101)$, and so on, 7 states can be obtained. State transition diagram is drawn, as shown in Fig. 8. Similarly, the state transition diagram of 3-order Galois LFSR in Fig. 7 is drawn, as shown in Fig. 9.

Conventional APUF circuits generally use an LFSR as an challenge extension circuit. When the initial challenge (seed) is input into the APUF circuit to produce a response bit, LFSR start to shift from the initial challenge (seed), and then the shifted challenge is input to the APUF circuit to produce a response bit, and so on. In essence, LFSR is used as a pseudo-random number generator to generate random challenges. The dual-LFSR APUF proposed in this paper uses two LFSRs to generate random challenges. We first input seed to the two LFSRs, and then shift the two LFSRs respectively under the same clock signal. Finally, At a certain clock, either of both

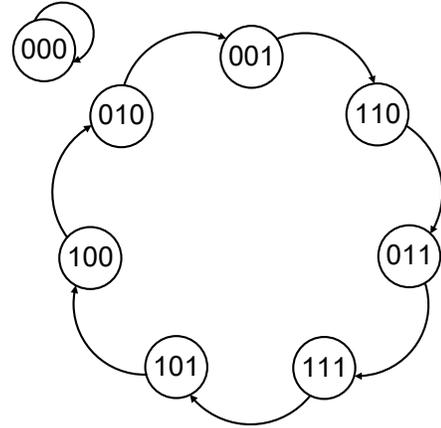
Fig. 8.  State transition diagram of 3-order LFSR in Fig.6

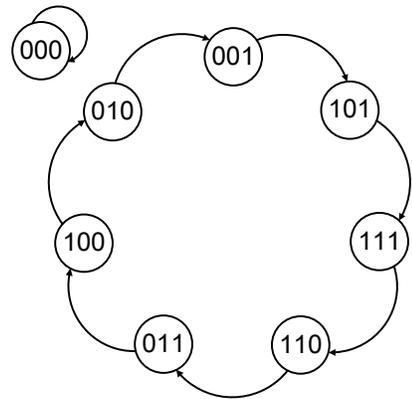
Fig. 9.  State transition diagram of 3-order LFSR in Fig.7

output sequences of the two LFSRs is selected as the challenge according to the previous response bit.

After the above analysis, the realization of dual LFSR structure needs to meet several conditions: 1) The useful sequences of the two LFSR should be the same. If the useful sequences of two LFSRs are different, it is assumed that the useful sequence of the LFSR1 contains sequence-A , but the useful sequence of LFSR2 does not contain sequence-A. In this case, if sequence-A is used as the seed input for two LFSR, the LFSR1 can work normally to produce useful sequence, while the LFSR2 can't work normally and may not produce useful sequence, which will lead to the dual LFSR can't normally produce challenge signal. We should try our best to avoid this situation. 2) The period of the two LFSR should be as large as possible, that is, the useful sequences of the two LFSR should be as many as possible. In fact, the real challenge of the input APUF is derived from the useful sequences of the two LFSRs. As a strong PUF, APUF has a large CRPs set, so the challenge set size of underlying APUF circuit should be as large as possible. 3) The number of useful sequences (periods) of the two LFSRs should be the same. If two LFSRs have different numbers of useful sequences, the two LFSRs must have different useful sequences. Suppose the number of useful sequences of LFSR1 is α, and the number of useful sequences of the second LFSR2 is β, where α>β. In this case, even if the β sequences in useful sequence of LFSR1 are the same as those in LFSR2, LFSR2 has at least (β-α) more useful sequences that LFSR2 dose not have.

According to the previous design, the schematic diagram of dual-LFSR APUF is generally divided into three parts, as shown in Fig. 10:

①-Challenge Generator Unit, it consists of dual-LFSR structure and a control module;

②-APUF Unit, a conventional APUF circuit;

③-Post Processing Unit, including path compensation module (for better randomness), arbiter, Voter Module (for better reliability) and XOR Module (for generating 1 bit response).

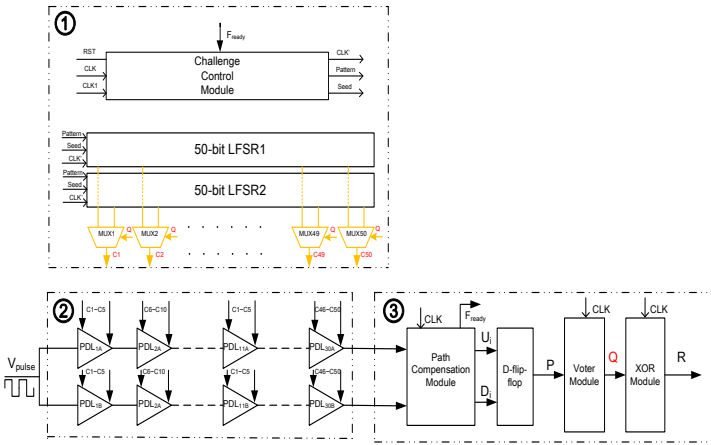
Fig. 10. Proposed schematic diagram of PUF

Based on the above analysis, it is necessary to construct two LFSRs of the same order to realize the dual-LFSR structure, at the same time, both LFSR can produce m-sequence. The cheme of response generation is as followings:

1. External challenge acts as the initial value (seed) of the two LFSRs. When the external challenges are input to dual LFSR structure, the two LFSRs begin to shift.
2. The real challenge of APUF circuit is determined by the previous response bit of the APUF. The current real challenge will be selected between the two LFSRs according to the previous response.
3. Assuming the APUF unit run 5 times, the 5 response bits generated by APUF are XORed as the final response bit of the dual-LFSR APUF.
4. Path Compensation Module, which compensates the shorter paths, is employed to improve the randomness. A Voter Module is used to improve the reliability of APUF.

Take 3-order LFSR as an example, and its state transition diagram is shown in Fig. 8 and Fig. 9 respectively. We know that the external challenges are the seeds of the two LFSRs. Assuming seed=001, the real challenge waveform generated by the dual LFSR is shown in Fig. 11. Fig. 11 is the waveform of each signal in the dual-LFSR structure when the external challenge(seed) is "001", where the response "R" is the assumed waveform of final response and the signal "Chg" is the internal real challenge waveform. As can be seen from the Fig. 11, which state value of LFSR is selected as the actual challenge "Chg" under the current clock is determined by the value of the

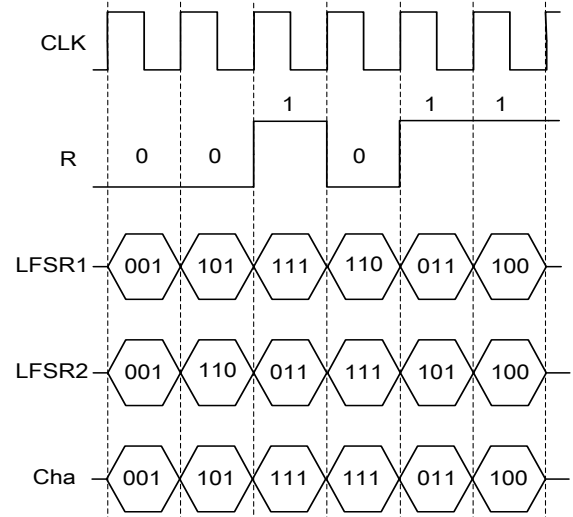
Fig. 11. Waveform of each signal in the dual-LFSR structure with seed "001"

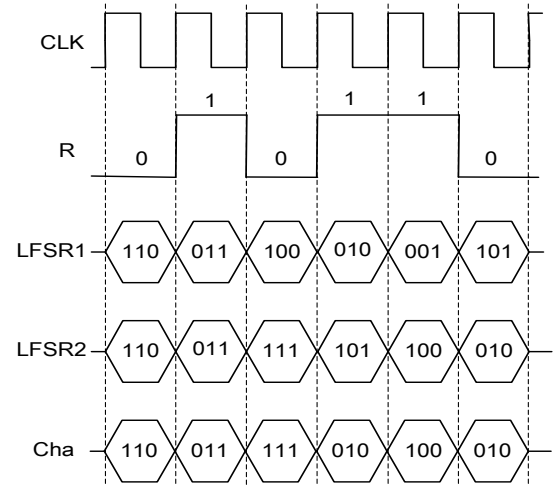
Fig. 12. Waveform of each signal in the dual-LFSR structure with seed "110"

response bit under the previous clock. As a contrast, the waveform of each signal in the dual-LFSR structure when the external challenge(seed) is "110" is as shown in Fig. 12.

Combined with Fig. 11 and Fig. 12, we can see that with the change of external challenge (seed), the jump mechanism of real challenge between the two LFSRs will also change greatly, and this time-variant obfuscation challenge method is difficult to decipher from the outside. However, how to ensure the stability of PUF while adopting the time-variant method of challenge obfuscation becomes an important problem.

The states of the main loop in the state transition diagram of two LFSRs should be identical(number of states and type of states), but the state jump rules of two LFSRs should be different. When the external challenge (seed) is fed into the dual LFSR structure, the seed does not fall outside the main cycle state of the two LFSRs. When the seed is input to both LFSRs, the states of two LFSRs in the following several shifts are uniquely determined by the seed. The reliability of the underlying APUF must be extremely high. When the seed is input to both LFSRs, the response bits of dual-LFSR APUF in the following several cycles are uniquely determined by the seed. In order to ensure the stability of PUF while adopting the time-variant method of challenge obfuscation becomes an important problem.

1. Two LFSRs can produce m-sequence;
2. The reliability of the underlying APUF is 1.0.

## III. PROPOSED AUTHENTICATION PROTOCOL

In a real authentication scenario, such as an RFID system, the attacker may eavesdrops the physical signals of a legitimate tag, captures and restores large number of interactive signals. Then, the attacker replays the exactly same signals to the reader according to the restored signals [10].

### A. The design principle of authentication against Man-in-the-Middle Attack

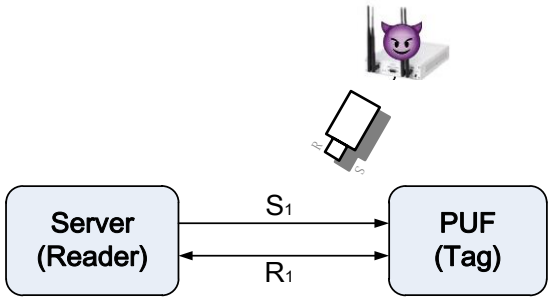

(a) Attackers eavesdrop on interaction signals

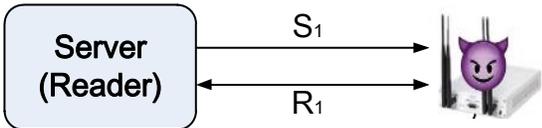

(b) The attacker passed the authentication by pretending to be legitimate

Fig.13. Attacker counterfeits legitimator for man-in-the-middle attack

During the authentication of a legitimate user, the server sends a certain challenge signal $S_1$ for the first time, and PUF replies to the corresponding response signal $R_1$. Assuming that these signals happen to be eavesdropped by the attacker, as shown in Fig. 13 (a). Assuming that the server is not aware of this, when the server sends the challenge signal for the second time, the attacker directly responds to the corresponding response signal according to the stored interaction signal to complete identity authentication, as shown in Fig. 13 (b).

In a man-in-the-middle attack, an attacker eavesdrops on and stores the physical signals of the interaction between the server and the PUF device, regardless of the content and meaning of the physical signals. At this level, an attacker can impersonate a legitimate device for authentication as long as the server repeatedly uses a certain challenge signal. To solve this problem, we consider whether information can be transmitted using certain parameter other than the physical signal. This paper takes high-frequency RFID as the authentication scenario, in which PUF is at the tag end and the server is at the reader end. In the specific authentication process, we load the information that needs to be interacted onto the carrier signal of the RFID system. In addition, during a authentication process, k-bit responses is required. In order to generate k-bit responses in PUF, we adopt the parallel structure of k dual-LFSR APUF and the mechanism of two authentications, as shown in Fig. 14.

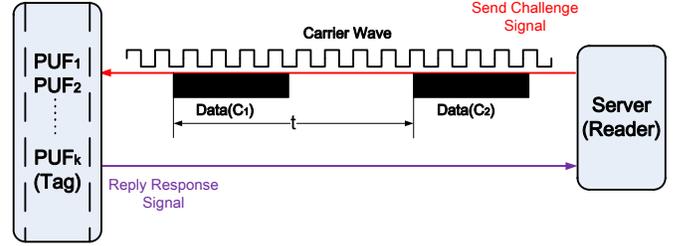

Fig.14. Authentication process in the parallel structure of k dual-LFSR APUF

During the authentication process, we input the same challenge signal to these k dual LFSR APUF. We load 2 groups of challenges onto the carrier in turn according to time, where the time interval between two successive groups of challenges is defined as t. Obviously, t is the information hidden in the interaction other than the physical signal. We can combine the challenges and t to send an challenge signal to the PUF device. In this way, even if the attacker successfully captures the challenge signal $S_2$ sent by the server and the response signal $R_2$ returned by the PUF device, it is impossible for the attacker to know the meaning of the time interval t. Assuming that the server sends the challenge signal $S_2$ again, the attacker will reply $R_2$ according to the stored $S_2$-$R_2$. Obviously, the attacker only collects the challenge signal $S_2$, but does not collect the hidden information t, or even does not know the meaning of the information t. However, for a legitimate PUF device, not only the hidden information t can be captured, but also its meaning can be analyzed and the response signal $R'_2$ can be returned after modifying the response signal accordingly, as shown in Fig. 15.

In order to realize the above scheme, we can extract the odd and even property of time interval t, and modify the response signal accordingly. As shown in Fig. 14, if the t is odd, LFSR will be selected according to the original method (LFSR1 if the previous response value is "0", otherwise LFSR2 will be selected) to generate the real challenge. If the t is even, LFSR will be selected to generate the real challenge according to the opposite method (LFSR2 if the previous response value is "0", otherwise LFSR1 will be selected). As shown in Fig. 5, the relationship between the current real challenge $\widehat{C}_t(k)$ and the previous response bit $R(k-1)$ is shown the following.

$$\widehat{C}_i(k) = \begin{cases} C_{1i}(k) & R(k-1) = 0 \\ C_{2i}(k), & R(k-1) = 1 \end{cases} \quad t-odd$$

$$\widehat{C}_i(k) = \begin{cases} C_{2i}(k) & R(k-1) = 0 \\ C_{1i}(k), & R(k-1) = 1 \end{cases} \quad t-even$$

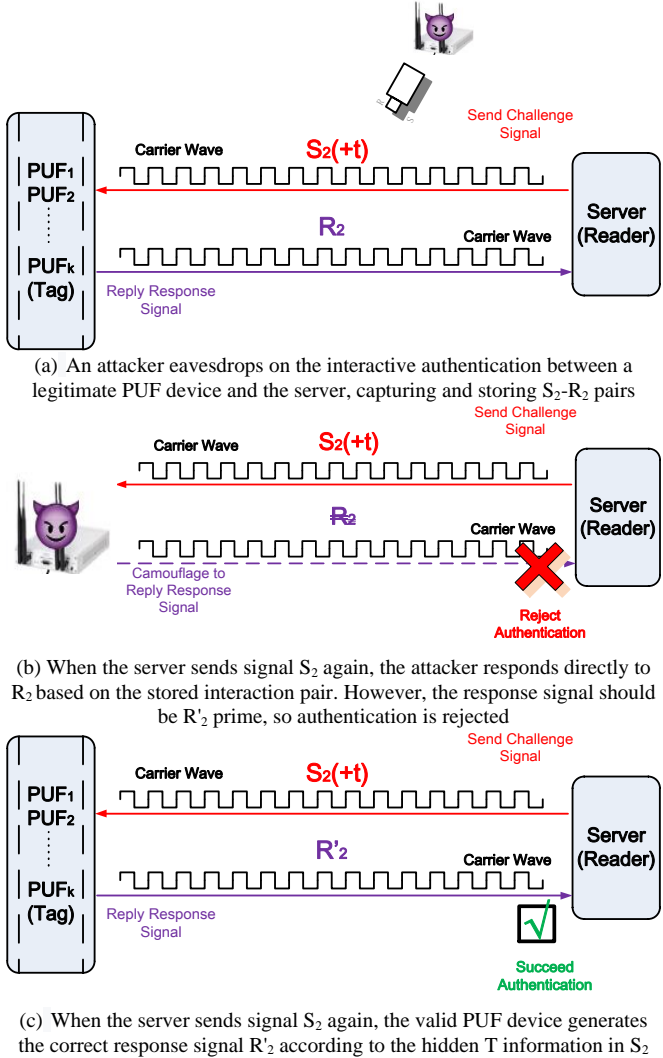

(a) An attacker eavesdrops on the interactive authentication between a legitimate PUF device and the server, capturing and storing $S_2$-$R_2$ pairs

(b) When the server sends signal $S_2$ again, the attacker responds directly to $R_2$ based on the stored interaction pair. However, the response signal should be $R'_2$ prime, so authentication is rejected

(c) When the server sends signal $S_2$ again, the valid PUF device generates the correct response signal $R'_2$ according to the hidden T information in $S_2$

Fig.15. Schematic of authentication against man-in-the-middle attack

### B. Authentication Device

Authentication devices generally include PUF device and server device. The PUF device is the label end. Note that the label end does not have a separate power supply, and the energy consumption of the label end comes from the electromagnetic signals emitted by the reader. The server device is the reader end. Note that the reader is connected to a larger software platform on which each module can be modeled.

The PUF device is implemented by k parallel APUF circuits and k matching parallel dual-LFSR challenge obfuscation circuits, challenge preprocessor and shift register circuits, the specific block diagram is shown in Fig. 16. As the front-end of PUF device, the challenge preprocessor is used to receive the signals sent by the server device, extract the challenge signals $C_1$ and $C_2$ of the data bits, and extract the time interval of the two groups of data bits. According to the parity of the time interval, a signal "M" is sent out to control the rotation mechanism of the two LFSRs in each dual-LFSR challenge obfuscation circuit. As the secondary challenge preprocessor, the dual-LFSR challenge obfuscation circuit receives N-bit challenge signal C sent by the challenge preprocessor as the seed of two LFSRs on the one hand, and receives "M" signals sent by the challenge preprocessor on the other hand, and switches the sequential selection of the two LFSRs according to the M value. Driven by two sets of input signals, k parallel dual-LFSR challenge obfuscation circuits generate N-bit obfuscated challenge C'. Note that the two LFSR structures of k dual-LFSR challenge obfuscation circuits are different to increase the complexity of recognition mechanism. As the secondary of dual-LFSR challenge obfuscation circuit, the challenge signal of APUF circuit has two sources, one is from the front dual-LFSR circuit, namely, obfuscated challenge C'. The other comes from the Disposable Interface, which is the challenge C directly from the external input. The Disposable Interface is mainly to extract the naked CRP of the underlying APUF circuit, and then establish a database of naked CRPs on the server side. Note that the Disposable Interface can only be used during the registration phase and is permanently fused after use to prevent an attacker from obtaining the naked CRPs through the interface. In addition, special attention should be paid to the preservation of PUF device before registration because the Disposable Interface is exposed after leaving the factory. It is better to adopt the physical shielding strategy to protect the Disposable Interface. As the secondary of K parallel APUF, the shift register can convert the parallel k-bit response $r_i$ (i=1,2······,k) into serial k-bit response $R_p$. In this way, the 2N-bit challenge signal $S_2(+t)$ can correspond one-to-one with the N-bit response signal $R_p$.

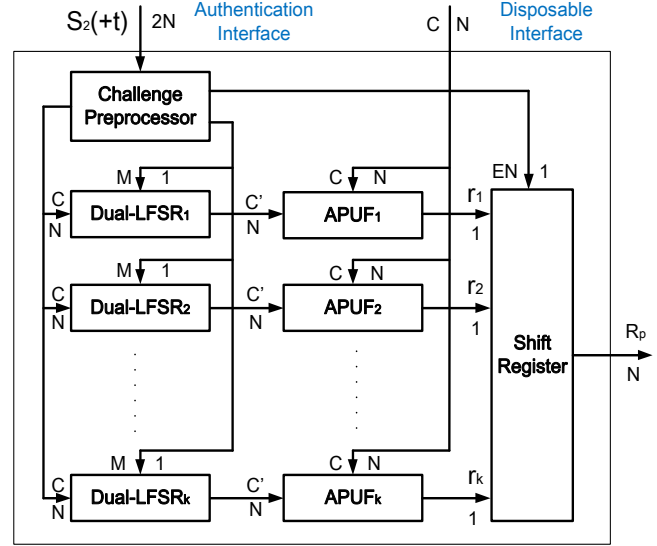

Fig.16. Components of PUF device

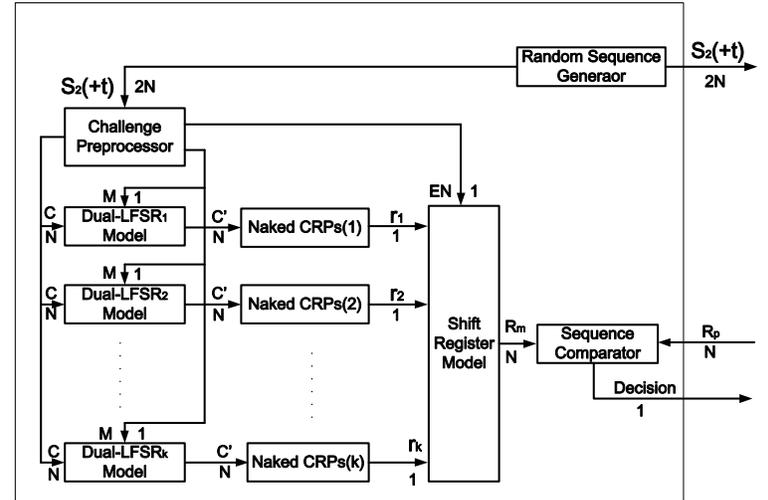

Fig.17. Components of sever

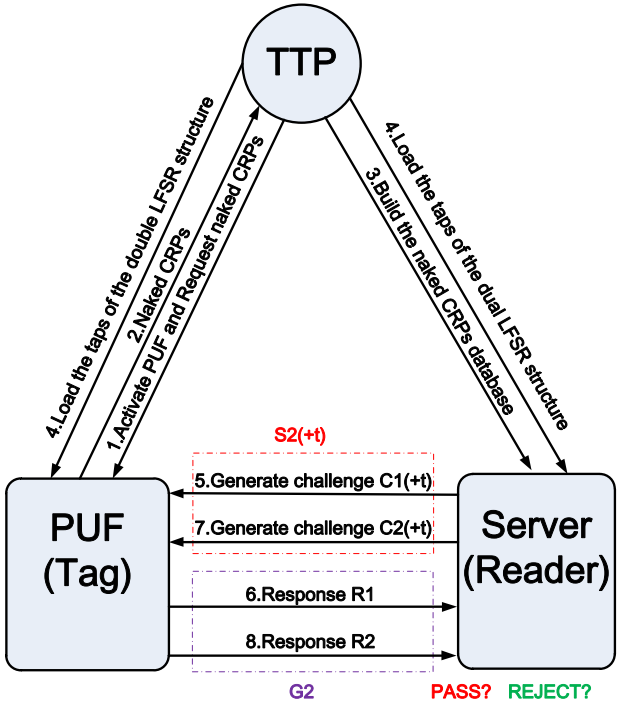

Fig.18. Authentication Protocol

Unlike PUF devices implemented on hardware platforms, server devices are implemented on software platforms. It mainly includes all naked CRPs of k parallel APUFs, k matching parallel dual-LFSR model, challenge preprocessor model, shift register model, random sequence generator and sequence comparator. The specific block diagram is shown in Fig. 17. Among them, all naked CRPs of k parallel APUFs have to be extremely reliable. Only the bottom circuit from $APUF_1$ to $APUF_k$ has high reliability, can ensure the dual-LFSR APUF circuit has high reliability, and then ensure the accuracy of the authentication process. Other modules such as k parallel dual-LFSR model, challenge preprocessor model and shift register model have exactly the same function as the corresponding module in Fig. 16, so it is easier to implement on the software platform. For the two extra modules, the random sequence module generates two N-bit random challenge $S_2$. Note that the time interval t between the two random challenges in the signal $S_2$ is also random. The Sequence Comparator is used to compare whether the N-bit response $R_m$ produced by the server device and the N-bit response $R_p$ produced by the PUF device are equal. If it is determined to be equal, the 1-bit decision value "1" will be generated and the authentication this time will be passed; if not, the 1-bit decision value "0" will be generated and the authentication this time will be refused. Pay attention to in the actual process of authentication, APUF model in server device can't be with APUF circuit in PUF device is exactly the same, and all the response bits of the APUF circuit in PUF device cannot be stable when the environment changes. We can set a threshold where $R_m$ and $R_p$ can be authenticated when there are only a few bits different. Note that the range of threshold must be strictly controlled by combining the size of k value, the reliability of APUF circuit and the accuracy of APUF model, otherwise false authentication may occur.

*C. Authentication Protocol*

The authentication protocol consists of three parts: PUF devices, server devices, and Trusted Third Party (TTP). There are two stages: registration stage and authentication stage. The authentication protocol is shown in Fig. 18.

The first phase is the registration stage. This phase relates to TTP and communication interaction with the server device and the communication interaction between TTP and PUF device. The first step is to activate the PUF device and then input an challenge through the Disposable Interface to the APUF circuit of the PUF device. In the second step, the APUF circuit of PUF device returns the corresponding response according to the input challenge of the first step. Through this step, CRPs of APUF circuit of k group of PUF device can be collected. After TTP has collected all CRPs, it is necessary to physically fuse the Disposable Interface to prevent attackers from obtaining naked CRPs of APUF circuit through the interface. The third step is to establish the naked CRPs database of k group APUF according to the collected k group CRPs set and transfer the built database to the server device. In the fourth step, The TTP builds models of other modules in the APUF device and transmits these models to the server device, based on the PUF device's parametric characteristics (mainly tap information) about the dual LFSRs provided by the manufacturer. At this point, the registration phase is complete. Note that the TTP is best served by the manufacturer so that it can be registered before leaving the factory, minimizing the scope for knowing the internal structure of the PUF device and preventing attackers from stealing naked CRPs through the Disposable Interface before registering the PUF device.

The second phase is the authentication stage. Authentication phase is the core of authentication protocol, which involves the communication interaction between server device and PUF device. In the first step of this stage, the server device needs to send the first set of random challenge $C_1$ to the PUF device and transfer the challenge internally to the challenge preprocessor. After this set of challenge is received by PUF device, PUF device obfuscates it, and then inputs k parallel APUF circuits to generate k-bit parallel response. After parallel serial processing, the final k-bit response signal $R_p$ is generated, and then sent it to the server. Similarly, at the server device end, random challenge $C_1$ received internally will also generate k-bit response signal $R_m$ after calculation by the internal module model. Finally, the sequence comparator in server end compares $R_p$ and $R_m$ and generates the value of the decision bit by determining whether $R_p$ and $R_m$ are equal. In the third step, the server device sends a second random challenge $C_2$ to the PUF device while passing the time interval t from within to the challenge preprocessor. Note in this step that the random sequence generator randomly determines the time interval t between the second $C_2$ and the first $C_1$ before the server sends the signal. In the fourth step, after the PUF device receives the signal, in addition to extracting and processing the random challenge $C_2$(specific as the second step), the t hidden between the two groups of challenges shall also be extracted, and the parity of the t value shall be judged to determine the "M" value, and then the k-bit response signal $R_p$ shall be generated. Similarly, the above operation will be directly repeated on the server side to generate the k-bit response signal $R_m$. The t value is simply sent directly to the challenge preprocessor, so there is no need to extract the size of the t value from the carrier. Finally, the server compares the $R_p$ and $R_m$ to produce the value of the decision bit.

Considerations during the authentication phase: 1) To ensure that no false authentication occurs, it is important to be very careful about determining the sequence comparator threshold size, which in principle should be as small as possible. 2) In the second authentication process, only if the decision position of the first and second authentication is "1", can the authentication be determined as passed. At the first authentication ($C_1$-$R_1$ interaction), the server device does not receive a response signal or an incorrect response signal, and the value of the decision bit is "0". At this point, the server will simply reject the authentication request, without the need for a second authentication.